\documentclass[12pt,a4paper]{article}

\usepackage{a4wide}
\usepackage{amssymb}
\usepackage{amsmath}
\usepackage{graphicx}
\usepackage{epsfig}
\usepackage{hyperref}
\usepackage{color}

\newcommand{\beq}{\begin{equation}}
\newcommand{\eeq}{\end{equation}}

\newcommand{\gbullet}{\!\bullet\!}

\def\[{\left[}
\def\]{\right]}
\def\({\left(}
\def\){\right)}
\def\[{\left[}

\begin{document}

\thispagestyle{empty}

\renewcommand{\thefootnote}{\fnsymbol{footnote}}
\setcounter{page}{1}
\setcounter{footnote}{0}
\setcounter{figure}{0}

\begin{center}
$$$$
{\Large\textbf{\mathversion{bold}
A note on defect Mellin amplitudes
}\par}

\vspace{1.0cm}

\textrm{Vasco Goncalves$^\text{\tiny 1}$, Georgios Itsios$^\text{\tiny 2}$}
\\ \vspace{1.2cm}
\footnotesize{\textit{
$^\text{\tiny 1}$ICTP South American Institute for Fundamental Research, IFT-UNESP,
S˜ao Paulo, SP Brazil 01440-070\\
$^\text{\tiny 2}$
Instituto de F\'i­sica Te\'orica, UNESP-Universidade Estadual Paulista, R. Dr. Bento T. Ferraz 271,
Bl. II, Sao Paulo 01140-070, SP, Brazil}  
\vspace{4mm}
}

\par\vspace{1.5cm}

\textbf{Abstract}\vspace{2mm}
\end{center}
We generalize the Mellin representation for a generic co-dimension flat defect CFT. We study the analytic structure of the Mellin amplitudes. We also compute Witten diagrams for a generic co-dimension flat defect CFT. 
\noindent

\setcounter{page}{1}
\renewcommand{\thefootnote}{\arabic{footnote}}
\setcounter{footnote}{0}

\setcounter{tocdepth}{2}

 \def\nref#1{{(\ref{#1})}}

\newpage

\parskip 5pt plus 1pt   \jot = 1.5ex

\section{Introduction}
Mellin amplitudes have proven to be a useful way to express correlation functions in conformal field theories. Its simple analytic structure together with the similarity to scattering amplitudes have made it an appealing representation. This was first noticed by Mack and Penedones in their seminal papers\cite{Mack,JPMellin}. A prototypical example is the computation of holographic correlators which turn out to have an extremely simple form when written in this language\cite{JPMellin}. A famous example is the graviton exchange between four minimally coupled massless scalars in $AdS_5$ which is given in terms of a few rational functions of the Mellin variables\cite{JPMellin}
\begin{align}
M(s_{ij})\propto \frac{6\gamma_{13}^2+2}{s_{13}-2}+\frac{8\gamma_{13}^2}{s_{13}-4}+\frac{\gamma_{13}-1}{s_{13}-6}-\frac{15}{4}s_{13}+\frac{55}{2}
\end{align}
where $s_{13},\gamma_{13}$ are the Mellin variables. The first striking feature of the answer is its simplicity, it amounts to a function with simple poles located at integers with the residues being at most a polynomial of degree $2$ in the other variable. In fact this  is not an accident it follows from the constraints that conformal symmetry imposes on Mellin amplitudes. For instance, poles of the Mellin amplitudes are associated with the operators that are exchanged in the OPE and the degree of the polynomial of the residue with the spin of the primary operator. This makes the Mellin language specially suitable to analyze holographic correlators where all higher spin single trace operators decouple from the spectrum. In fact one of the most recent applications of the Mellin formalism was the computation of holographic correlators bypassing the difficult and cumbersome task of the standard way to obtain the correlator. One of the main advantages of the method is that it only relies on general consistency conditions. This was first tried for four point functions in $\mathcal{N}=4$ at strong coupling\cite{Rastelli:2016nze,Rastelli:2017ymc,Zhou:2017zaw} and then to holographic four point functions in the $(2,0)$ theory\cite{Rastelli:2017udc}. 

Another recent application of Mellin amplitudes is the conformal bootstrap in Mellin space\cite{Gopakumar:2016wkt}. This is a particularly good example of an old idea that only was made efficient when the problem was expressed in the proper language. One of the most remarkable results of this approach is the determination of high order in $\epsilon$ of both anomalous dimension and OPE coefficients in Wilson-Fisher type of models. 

The Mellin formalism was recently applied to boundary conformal field theories. These have less symmetries but the Mellin formalism can still be applied and it turns out that Witten diagrams can be written in a simpler form in terms of Mellin variables. The boundary case can be thought as conformal field theory that lives only on half space, say $x_d>0$. The boundary setting is a special case of a defect CFT where the symmetry group is $SO(p + 1, 1) \times SO(q)$ which compares with $SO(d+1,1)$ of a conformal field theory(the co-dimension one defect $q=1$ corresponds to the boundary).  The present paper generalizes the Mellin formalism to a generic flat defect CFT. More precisely we will study: implications to the Mellin amplitude of both defect and bulk operator product expansions as well as Witten diagrams in a defect setting. 

Defects are particularly interesting to study both from the theoretical and practical point of view. On the theoretical part there are more degrees of freedom  present in the a defect CFT which allows to probe different physics. For example, it has been recently discovered the connection between defect CFT's and R\'enyi entropy\cite{Bianchi:2015liz}. On the practical side there is the advantage that the first nontrivial function that appears is the two-point function of bulk operators and this compares with the four point function of the non defect CFT. The boundary case is even more special in this respect since there are less cross ratios and the analysis of any bootstrap method is potentially simpler\footnote{This is in analogy with the recent conformal bootstrap studies of one dimensional CFTs where by construction the correlation functions have less conformal cross ratios.}.

\section{Defect Mellin amplitudes}
Symmetry places big constraints on the structure of correlation functions of local operators. A well known example is how conformal symmetry fixes the form of two- and three-point correlators up to some constants in a conformal field theory. A defect CFT has less symmetries but still they are strong enough to fix the form of one-point function of a bulk operator as well as the correlator between one bulk operator and a defect one\cite{Billo:2016cpy}
\begin{align}
\langle\mathcal{O}(P) \rangle =\frac{a_{\mathcal{O}}}{(P\circ P)^{\Delta/2}} \, , \, \ \ \ \  \langle\mathcal{O}(P_1) \hat{\mathcal{O}}(\hat{P}_2)\rangle=\frac{b_{\mathcal{O}\hat{\mathcal{O}}}}{(P_{1}\circ P_1)^{\frac{\Delta_{\mathcal{O}}-\Delta_{\hat{\mathcal{O}}}}{2}}(P_1\gbullet P_2)^{\Delta_{\hat{\mathcal{O}}}}} \, ,
\end{align}
where the hat stands for operators living on the defect and where we used the defect embedding formalism (see \cite{Billo:2016cpy,Gadde:2016fbj,Fukuda:2017cup} for details or appendix A). Some of these coefficients are the same as in the CFT without a defect. However the dimension of defect operators, the bulk to defect coupling and one-point function coefficient are new sets of conformal data.  One open question is if one can constrain the allowed space of the defect data like it has been for the CFT case \cite{Rattazzi:2008pe}. This is more nontrivial for the defect setting since one of the OPE channels is not positive definite\footnote{Numerical bootstrap was attempted for the boundary case in \cite{arXiv:1210.4258,arXiv:1502.07217} where it was assumed positivity in both OPE channels.}. 
 
The simplest non-trivial correlation function in a defect CFT is the two-point function of two bulk operators  which is fixed up to a non-trivial function of two cross ratios for generic co-dimension and one cross ratio for co-dimension one defects\cite{Billo:2016cpy}
\begin{align}
& \langle \mathcal{O}_{\Delta_1}(P_1)\mathcal{O}_{\Delta_2}(P_2) \rangle_q = \frac{\mathcal{A}(\xi,\cos \phi)}{(P_1\circ P_1)^{\Delta_1/2}(P_2\circ P_2)^{\Delta_2/2}} \, , \label{eq:twopointfunction}
\\[10pt]
& \xi= \frac{-2P_1\cdot P_2}{(P_1\circ P_1)^{1/2} \, (P_2\circ P_2)^{1/2}},\, \ \ \ \cos \phi = \frac{P_1\circ P_2}{(P_1\circ P_1)^{1/2} \, (P_2 \circ P_2)^{1/2}} \, .
\end{align}
For co-dimension one defects $\cos \phi =1$ and so the two-point function depends only on one cross ratio.

Higher correlation function with $n$ bulk and $m$ defect operators will depend on more cross ratios. These can be constructed with the following scalar products
\begin{align}
-2P_i\cdot P_j =P_{ij},\, \ \ P_i\circ P_i,\, \ \ \ P_i\circ P_j,\, \ \ \ -2P_i\cdot \hat{P}_J=P_{iJ},\, \ \ \ -2\hat{P}_I\cdot \hat{P}_J=\hat{P}_{IJ}.
\end{align}
Symmetry fixes correlators in terms of a non-trivial function of the cross ratios. However the existence of the operator product expansion constraints further the analytic structure of these non-trivial functions. The analytic structure turns out to be simpler when it is expressed in terms of its Mellin transform, very much like what happens with Fourier space for scattering amplitudes in quantum field theory. Let us define the Mellin amplitude, $M$,  of an $n$ bulk and $m$ defect operator correlation function  as
\begin{align}
\langle \mathcal{O}_{1}\dots \mathcal{O}_n \hat{\mathcal{O}_1}\dots \hat{\mathcal{O}}_m\rangle = \int [d\delta_{ij}][d\alpha_i][d\nu_{ij}]&[d\gamma_{iI}][d\beta_{IJ}]\,M(\delta,\alpha,\nu,\gamma,\beta)\nonumber\\
&P_{ij}^{-\delta_{ij}}(P_{i}\circ P_i)^{-\alpha_i/2}(P_i\circ P_j)^{-\nu_{ij}}P_{iI}^{-\gamma_{iI}}P_{IJ}^{-\beta_{IJ}} \, ,
\end{align}
where the integration contours runs parallel to the imaginary axis. The integration variables are not all independent as they satisfy 
\begin{equation}
 \begin{aligned}
& \sum_{j \ne i}(\delta_{ij}+\nu_{ij})+\sum_{I}\gamma_{iI}+\alpha_i=\Delta_i\,,\, \ \ \ \ \sum_{J \ne I}\beta_{IJ}+\sum_{i}\gamma_{iI}=\hat{\Delta}_I\label{eq:ConstraintsMellin} \, ,
\\
& \delta_{ij} = \delta_{ji} \, \qquad \nu_{ij} = \nu_{ji} \, , \qquad \beta_{IJ} = \beta_{JI} \, .
 \end{aligned}
\end{equation}
These conditions guarantee the correct scaling of both bulk and defect operators. The defect not only possess an operator product expansion when a subset of operators is close together\footnote{Let us remark that this property is valid for both bulk or defect operators.} but also when a subset of bulk operators is close to the defect. Both properties  can be made precise in the context of radial quantization around one-point in the bulk or one-point on the defect, respectively. In the first case one has a subset of $k$ operators that can be replaced by a family of primary operators and their descendants placed on the point where one is doing radial quantization around\cite{NaturalMellin}. There is just one subtlety compared to the CFT case, that is now one is not allowed to choose this point to be the origin since the defect is placed there. So take the point of the radial quantization to be random bulk point labelled by $P_0$
\begin{align}
\mathcal{O}_1(P_1)\dots \mathcal{O}_k(P_k) =\sum_p  \sum_{m=0}C_{p,\nu_1\dots \nu_l}^{\mu_1\dots \mu_m}(P_1,\dots,P_k,P_0)\partial_{\mu_1}\dots \partial_{\mu_m}\mathcal{O}_p^{\mu_1\dots\nu_l}(P_0)\label{eq:ManyOPE} \, .
\end{align}
This is an operator equation that is valid as long as it is possible to find a ball centered around $P_0$ that encircles all this subset of operators and leaves outside the other subset. The action of the dilatation operator on the subset of $k$ operators has the effect of changing the positions  $P_i\rightarrow e^{-\lambda}(P_i-P_0)+P_0$ on the left hand. The effect on the right hand side is given by dressing each term by a factor $e^{-\lambda(\Delta_p+m-\sum_i\Delta_i)}$. 

It is also possible to express a subset of $k$ operators in terms of defect degrees of freedom. This can be done by performing radial quantization around a defect point $\hat{P}$\cite{Billo:2016cpy}
\begin{align}
\mathcal{O}_1(P_1)\dots \mathcal{O}_k(P_k)=\sum_{\hat{p}}C_{\hat{p},\nu_1\dots \nu_l}(P_1,\dots,P_k)\hat{\mathcal{O}}_{\hat{p}}^{\mu_1\dots\nu_l}(\hat{P})\label{eq:defectOPE} \, ,
\end{align}
where the indices are perpendicular to the defect. One of the most important differences compared to the bulk OPE is that this equation is valid even with just one operator on the left hand side. We will analyze in the following this specific case since it has the advantage that it does not have mixed effects with the bulk OPE. 
The action of the dilatation operator by a factor $e^{-\lambda}$ on all operators on the left hand side has the effect of dressing each term in the sum on the right hand side by $e^{-\lambda(\hat{\Delta}_p-\Delta_\mathcal{O})}$.  

One of the advantages of Mellin amplitudes is that it makes simple the action of the dilatation operator on a subset of operators\cite{NaturalMellin}. For example,  the effect of doing the dilatation operation around the bulk point $P_0$ to a subset of $k$ operators is 
\begin{align}
&\int [dQ]\,M(\delta,\alpha,\nu,\gamma,\beta)e^{2 \, \lambda \sum_{i<j}^k \delta_{ij}}\prod_{i<j}^{k}P_{ij}^{-\delta_{ij}}\nonumber\\
&\prod_{k<i<j}P_{ij}^{-\delta_{ij}}(P_i\circ P_j)^{-\nu_{ij}}\prod_{k<i}(P_{i}\circ P_i)^{-\alpha_i/2}\prod_{k<i,I}P_{iI}^{-\gamma_{iI}}P_{IJ}^{-\beta_{IJ}}\sum_{q=0}^{\infty}e^{-q\lambda}R_q(P_i,\hat{P}_I) \, ,
\end{align}
where $[dQ]$ was used to denote all integration variables. The last factor with $R$ comes from taking the large $\lambda$ of crossed terms
\begin{align}
\sum_{q=0}^{\infty}e^{-q\lambda}R_q(P_i,\hat{P}_I)=&\prod_{i< j<k} \big( ((x_{i,\perp}-x_{0,\perp})e^{-\lambda}+x_{0,\perp})\cdot ((x_{j,\perp}-x_{0,\perp})e^{-\lambda}+x_{0,\perp}) \big)^{-\nu_{ij}}\\
&\prod_{i<k<j}(x_j^2-2(x_i-x_0)\cdot x_je^{-\lambda}+e^{-2\lambda}(x_i-x_0)^2+x_0^2-2x_j\cdot x_0)^{-\delta_{ij}}\nonumber\\
&\prod(x_I^2-2(x_i-x_0)\cdot x_Ie^{-\lambda}+e^{-2\lambda}(x_i-x_0)^2+x_0^2-2x_0\cdot x_I )^{-\gamma_{iI}}\nonumber\\
&\prod_{i<k} \big( ((x_{i0,\perp})e^{-\lambda}+x_{0,\perp})\cdot ((x_{i0,\perp}e^{-\lambda}+x_{0,\perp}) \big)^{-\alpha_{i}}\nonumber.
\end{align}
By definition $R_q$ is an homogeneous polynomial of degree $q$ in $x_i$ and related with the exchange of a spin $q$ operator in the OPE. Consistency with (\ref{eq:ManyOPE}) implies that the Mellin amplitude has simple poles at
\begin{align}
\sum_i\Delta_i-2\sum_{i<j}^{k}\delta_{ij}=\Delta_p-l_p+m \, ,
\end{align}
where $l_p$ is the spin of the primary operator with scaling dimension $\Delta_p$.

The existence of a defect OPE constrains even further the analytic structure of the Mellin amplitude as can be seen by applying (\ref{eq:defectOPE}) in the specific case of just one operator
\begin{align}
\mathcal{O}(P_1)=\sum_{\hat{p}}C_{\hat{p},\nu_1\dots \nu_l}(P_1)\hat{\mathcal{O}}_{\hat{p}}^{\mu_1\dots\nu_l}(\hat{P}).
\end{align}
This implies that the Mellin amplitude has poles at the positions 
\begin{align}
\Delta_1-(\alpha_1+\sum_{j \ne 1}\nu_{1j})=\hat{\Delta}_p-\hat{l}_p+m \, , \label{eq:polesdefectchannel}
\end{align}
where $\hat{l}_p$ is the transverse spin. The presence of the spin can be seen by expanding $e^{-\lambda}x_{1,\perp}$ in  $P_{1j}$ and $P_{1I}$  and noticing that the numerator is a polynomial in $x_{1,\perp}$.

Another nice property of Mellin amplitudes is that they possess similar factorization properties that scattering amplitudes have. By this we mean that the residues of the poles are naturally divided into lower point correlation functions. In the CFT case this statement can be made very explicit by using the shadow formalism\cite{NaturalMellin,Goncalves:2014rfa}, however this formalism is not well developed in the defect setting(see \cite{Fukuda:2017cup} for progress in these directions). We will not try to solve this problem here, instead in the following we will focus on the two-point function of bulk operators since this is one of the simplest and most important case for applications\footnote{Let us remark that once one obtains that one only needs the residue for the $m=0$, since the action of the Casimir equation on the Mellin amplitude generates a recurrence relation, that gives higher values of $m$\cite{NaturalMellin,Costa:2012cb,Goncalves:2014rfa}. }.

The Mellin amplitude associated with the two-point function (\ref{eq:twopointfunction})
\begin{align}
& \mathcal{A}(\xi,\cos \phi)= \int [d\delta_{12}][d\nu_{12}]M(\delta_{12},\nu_{12})\xi^{-\delta_{12}}(\cos\phi)^{-\nu_{12}},
\end{align}
where we have solved the constraints (\ref{eq:ConstraintsMellin}). For co-dimension one defects there is just one integration variable, $\delta_{12}$ since $x_{1,\perp}\cdot x_{2,\perp}$ can be expressed in terms of both $x_{1,\perp}^2$ and $x_{2,\perp}^2$. The two-point function amplitude can be decomposed in bulk conformal blocks 
\begin{align}
\mathcal{A}(\xi,\cos \phi) = \sum_p c_{\Delta_1\Delta_2\Delta_p}a_{\Delta_p} G_{\Delta_p,l_p}(\xi,\cos \phi) \, ,
\end{align}
where $G_{\Delta_p,l_p}$ is a bulk conformal block that satisfies the Casimir equation $\mathcal{D}_{bulk}G_{\Delta,l}=0$ with
\begin{align}
&\mathcal{D}_{\textrm{bulk}} \equiv 2\xi^2\left(2+\xi\cos\phi+2\cos^2\phi\right)\frac{\partial^2}{\partial \xi^2}+2\sin^2\phi \left(2\sin^2\phi-\xi\cos\phi\right)\frac{\partial^2}{\partial \cos\phi^2} \label{eq:bulkCasimir} \\
&-4\xi \sin^2\phi\left(\xi +2\cos\phi\right)\frac{\partial^2}{\partial \xi\partial\cos\phi}+2\xi\left[2(1+\cos^2\phi)-(2d-\xi\cos\phi)\right]\frac{\partial}{\partial \xi}\nonumber\\
&+\left[2\xi(q-2+\cos^2\phi)-4\cos\phi \sin^2\phi\right]\frac{\partial}{\partial \cos\phi}
-\left[\Delta_{12}^2 \cos \phi  \left(\cos \phi + \frac{\xi}{2} \right) -\Delta_{12}^2+2C_{\Delta_k,J} \right] \, ,\nonumber
\end{align}
with $C_{\Delta_k,J}$ being the eigenvalue of the Casimir operator. The conformal block takes into account the contribution of a primary operator with dimension $\Delta_p$ and spin $l_p$ and $c_{\Delta_1\Delta_2\Delta_p}$ is the usual CFT OPE coefficient. Conformal blocks admit an expansion in powers of $\xi^{\frac{\Delta_p-l_p}{2}+m}$ with positive integer $m$\cite{Billo:2016cpy}. This expansion is associated with the light cone limit of the two bulk operators. A direct consequence of this is that the Mellin amplitude should have poles located at $2\delta_{12}=\Delta_p-l_p+m$.

The function multiplying the $m=0$ term in the bulk conformal block is independent of the co-dimension and is given by an hypergeometric function\cite{Billo:2016cpy}
\begin{align}
G_{\Delta_p,l_p}=\xi^{\frac{\Delta_p-l_p}{2}}\Bigg[\underbrace{\sin^{l_p}\phi \,\!_2F_1\!\left(\frac{l_p+\Delta_p}{4},\frac{\Delta_p+l_p}{4},\frac{\Delta_p+l_p+1}{2},\sin^2\phi\right)}_{g_0(\phi)}+O(\xi)\Bigg].
\end{align}
Finding the residue for the $m=0$ is now a straightfoward task since it can now be mapped to the same problem that was solved in the CFT case\cite{Costa:2012cb}
\begin{align}
g_0 =&\frac{\Gamma\left(\frac{\Delta+J+1}{2}\right)\left(\frac{\Delta-1}{2}\right)_{J/2}}{2^{J-1}\Gamma^2\left(\frac{\Delta+J}{4}\right)\Gamma^2\left(\frac{\Delta+J+2}{4}\right)} \int \frac{d\gamma_{12}}{8\pi i}(\cos \phi)^{\frac{J-\Delta-1-2\gamma_{12}}{2}} Q_{J,0}(\gamma_{12})\\[5pt]
&\Gamma^2\Big(-\frac{2\gamma_{12}+1}{4}\Big)\Gamma\Big(\frac{\Delta+1-J+2\gamma_{12}}{4}\Big)\Gamma\Big(\frac{\Delta+3-J+2\gamma_{12}}{4}\Big) \, , \nonumber
\end{align}
where the function $Q_{J,0}(\gamma_{12})$ is a polynomial of degree $J/2$ in $\gamma_{12}$
\begin{align}
Q_{J,0}(\gamma_{12})= &\frac{2^{J/2}\left(\frac{\Delta-J}{4}\right)_{J/2}\left(\frac{\Delta+2-J}{4}\right)_{J/2}}{\left(\frac{\Delta-1}{2}\right)_{J/2}}\times\\
&\times\,\!_3F_2\left(-\frac{J}{2},\frac{\Delta-1}{2},-\frac{2\gamma_{12}+1}{4};\frac{2+\Delta-J}{4},\frac{\Delta-J}{4},1\right).
\end{align}
The solution of the bulk conformal blocks can be written in terms of sums of the elementary block $g_0$, so it is possible to write the residue of the pole for the case of $m>0$ in terms of sums of $Q_{J,0}(\gamma_{12})$\footnote{Alternatively one can act with the Casimir equation (\ref{eq:bulkCasimir}) and this generates a recurrence relation for $Q_{J,m}(\gamma_{12})$.}. There is a natural scalar product such that the product of two functions $Q$ with different $J$ is orthogonal\cite{Costa:2012cb,Goncalves:2014ffa}
\begin{align}
&\int \frac{d\gamma_{12}}{4\pi i}Q_{J,0}(\gamma_{12})Q_{J',0}(\gamma_{12})\Gamma^2 \!\left(-\frac{2\gamma_{12}+1}{4}\right) \Gamma\! \left(\frac{2 \gamma_{12}+\Delta+1-J}{4}\right) \Gamma\! \left(\frac{2 \gamma_{12}+\Delta+3-J}{4}\right)\nonumber\\
&= \delta_{J,J'} \, \frac{(-1)^{J/2} \, 2^{J+1} \, \big(   \frac{J}{2}  \big)!}{\big(   \frac{\Delta - 1}{2}  \big)^2_{J/2}} \, \frac{\Gamma^2\big(   \frac{\Delta + J}{4}  \big) \, \Gamma^2\big(   \frac{\Delta + J + 2}{4} \big)}{\big(  \Delta + J - 1   \big) \, \Gamma\big(   \frac{\Delta - 1}{2}  \big)} \, .
\end{align}
Let us remark that this property of the functions $Q_{J,0}$ played an important role in the context of the conformal bootstrap in the Mellin space\cite{Gopakumar:2016cpb,Gopakumar:2016wkt}. Another application is to extract corrections to the dimension and OPE coefficients due to the exchange of an operator in another channel(see \cite{Costa:2014kfa} for examples of this in the context of CFTs).

The two-point function amplitude $\mathcal{A}$ can also be decomposed in terms of defect conformal blocks 
\begin{align}
\mathcal{A}(\xi,\cos \phi) = \sum_{\hat{\mathcal{O}}} b_{1\hat{\mathcal{O}}}b_{1\hat{\mathcal{O}}} G_{\hat{\Delta},s}(\xi,\cos \phi) \, .
\end{align} 
These are eigenfunctions of the differential operators 
\begin{align}\label{casimdef}
\mathcal{D}_{\textrm{def}}^{L^2}\mbox{ }\widehat{f}_{\widehat{\Delta},0,s}(\xi,\phi)=0, \qquad \mathcal{D}_{\textrm{def}}^{S^2}\mbox{ }\widehat{f}_{\widehat{\Delta},0,s}(\xi,\phi)=0 \, ,
\end{align}
that are defined as  
\begin{align}
&\mathcal{D}^{S^2}_{\textrm{def}} \equiv 4\cos\phi(1-\cos\phi) \frac{\partial^2}{\partial \cos\phi^2}+2(1-q\cos\phi) \frac{\partial}{\partial\cos\phi}\nonumber\\
&+16\cos\phi(1-\cos\phi)\frac{\partial^2}{\partial \xi^2}-16\cos\phi(1-\cos\phi)\frac{\partial^2}{\partial \xi\partial\cos\phi}-4(1-q\cos\phi)\frac{\partial}{\partial \xi}+\widehat{C}_{0,s},\\[5pt]
&\mathcal{D}^{L^2}_{\textrm{def}} \equiv (4-(\xi+2\cos\phi)^2) \frac{\partial^2}{\partial \xi^2} - (p+1)(\xi+2\cos\phi) \frac{\partial}{\partial \xi}+\widehat{C}_{\Delta,0}.
\end{align}
Again $\widehat{C}_{0,s}$ and $\widehat{C}_{\Delta,0}$ are the corresponding eigenvalues. The action of the differential operators can be written in a simpler way in the variable $\chi=\xi+2\cos\phi$. However we will keep using $\xi$ to make it easier to compare with bulk OPE. 

The Mellin amplitude also has poles corresponding to the exchange of defect operators. Their location is given by (\ref{eq:polesdefectchannel})
\begin{align}
l - m - \hat{\Delta} + \delta_{12}=0
\end{align}
for the case of a two-point correlation function of bulk operators. Notice that the sign of $\delta_{12}$ is opposite as compared to the bulk OPE pole, this is just a consequence of the fact that for bulk OPE one has $\xi\rightarrow 0$ and for defect OPE one has the opposite limit $\xi\rightarrow \infty$.  What remains to be done is to determine the residue of the pole. Recall that the solution of the differential equations is given by 
\begin{align}
G_{\hat{\Delta},0}=&\alpha_{s,q}(\xi+2\cos\phi)^{-\hat{\Delta}}\!_2F_1\Big(\frac{q+s-2}{2},-\frac{s}{2},\frac{q-1}{2},\sin^2\phi\Big)\times\nonumber\\
&\!_2F_1\Bigg(\frac{\hat{\Delta}+1}{2},\frac{\hat{\Delta}}{2},\hat{\Delta}+1-\frac{p}{2},\frac{4}{(\xi+2\cos\phi)^2}\Bigg) \, ,
\end{align}
where $\alpha_{s,q}$ is a normalization constant. The residues could be read by expanding the defect conformal block for large $\xi$. The leading term is given just by
\begin{align}
\!_2F_1\Big(\frac{q+s-2}{2},-\frac{s}{2},\frac{q-1}{2},\sin^2\phi\Big) \, ,
\end{align}
which is a polynomial of $\cos\phi$ for physical values of the spin $s$. So, its Mellin transform is not well defined.

\section{Witten diagrams}
One  of the main applications of Mellin amplitudes has been the computation of SUGRA correlators. These correlators are constructed with Witten diagrams. This section deals with the computation of Witten diagrams for a generic co-dimension defect.  The setup is a generalization of \cite{Rastelli:2017ecj} where the holographic correlator in $AdS_{1+d}$ has a preferred  $AdS_{1+d-q}$ subspace that gets integrated. The effective action considered here is 
\begin{align}
S=\int_{AdS_{1+d}} \mathcal{L}_{\textrm{bulk}}[\Phi_i]+\int_{AdS_{1+d-q}} \Big(\mathcal{L}_{\textrm{defect}}[\phi_{I}]+\mathcal{L}_{\textrm{int}}[\Phi_i,\phi_I] \Big) \, ,
\end{align}
where $\Phi$ and $\phi$ are fields that live on $AdS_{1+d}$ and $AdS_{1+d-q}$ respectively. This serves also as an example of Mellin amplitudes since these correlators are naturally expressed in terms of this language. 

Contact or exchange Witten diagrams have their origin in different terms in the effective action. An interaction term of the form $\Phi_{i_1}\dots\Phi_{i_n}\phi_{I_1}\dots \phi_{I_m}$ gives rise to a contact Witten diagram with $n$ bulk operators and $m$ defect ones
\begin{align}
W_{n,m}(P_i,\hat{P}_I)&=\int_{AdS_{1+d-q}}dW\prod_{i,I}G_{B\partial}^{\Delta_i}(P_i,W)G_{B\partial}(\hat{P}_I,W)\label{eq:AdSCorrelatorGeneric},\\
G_{B\partial}^{\Delta}(P,W)&= \frac{1}{(-2P\cdot W)^{\Delta}}.\label{eq:bulktoboundarypropagator}
\end{align}
For $n=1,m=0,1$ the diagrams have no cross ratio as can be seen by explicitly computing them. All the steps are essentially the same as in the boundary setting \cite{Rastelli:2017ecj}, so only the final result is shown
\begin{align}
&W_{1,0}(P) = \frac{\pi^{\frac{d-q}{2}}\Gamma\left(\frac{\Delta}{2}\right)\Gamma\left(\frac{\Delta+q-d}{2}\right)}{2\Gamma(\Delta)}\frac{1}{(P\circ P)^{\frac{\Delta}{2}}} \, ,
\\[5pt]
&W_{1,1}(P_1,\hat{P}_2) = \frac{\pi^{\frac{d-q}{2}}\Gamma\left(\frac{\Delta_1+\hat{\Delta}_2-d+q}{2}\right)\Gamma\left(\frac{\Delta_1-\hat{\Delta}_2}{2}\right)}{2\Gamma(\Delta_1)}\frac{1}{(P_1\circ P_1)^{\frac{\Delta_1-\hat{\Delta}_2}{2}}(-2P_1\cdot \hat{P}_2)^{\hat{\Delta}_2}} \, .
\end{align}
On the other hand the case of two bulk operators, {\textit{i.e}} $n=2,m=0$, depends on two cross ratios.  The introduction of Schwinger parameters in (\ref{eq:AdSCorrelatorGeneric}) for each propagator makes trivial the integration over $AdS_{d+1-q}$. This leads to
\begin{align}
W_{2,0}(P_1,P_2)=&\frac{\pi^{\frac{d-q}{2}}\Gamma\big(\frac{\Delta_1+\Delta_2+q-d}{2}\big)}{2 \, \Gamma(\Delta_1)\Gamma(\Delta_2)}\int \frac{d\rho \, d\alpha_1\, d\alpha_2\,\alpha_1^{\Delta_1}\alpha_2^{\Delta_2}\rho^{\frac{\Delta_1+\Delta_2}{2}}}{\rho\alpha_1\alpha_2}
\nonumber\\
&\times\delta(1-\alpha_1-\alpha_2)e^{
-\rho \left(\alpha_1^2\,x_{1,\perp}^2+\alpha_2^2\,x_{2,\perp}^2 \right)} e^{-\rho\alpha_1\alpha_2 \left(2 x_{1,\perp}\cdot\,x_{2,\perp}+(x_1-x_2)^2 \right)} \, ,
\end{align} 
where we have used  $1=\int_0^{\infty}d\rho \delta(\rho-\alpha_1-\alpha_2)$, rescaled $\alpha_1\rightarrow \rho\,\alpha_1,\,\alpha_2\rightarrow \rho \alpha_2$ and the integral $\alpha_i$ range from $0$ to $1$. Now we represent $e^{-z}$ with its inverse Mellin transform 
\begin{align}
e^{-z}=\int_{-i\infty }^{i\infty}\Gamma(\tau) z^{-\tau}d\tau
\end{align}
for the exponentials $-\rho \, \alpha_1\alpha_2(x_1-x_2)^2$ and $-2\rho \, \alpha_1\alpha_2 \, x_{1,\perp}\cdot x_{2,\perp}$. Notice that the co-dimension one defect CFT is special since $(\alpha_1 x_{1,\perp}+\alpha_2 x_{2,\perp})$ is just a number, so it is not necessary to analyze the factors $\alpha_1\alpha_2 \, x_{1,\perp}\cdot x_{2,\perp}$ and $\alpha_1^2x_{1,\perp}^2+\alpha_2^2x_{2,\perp}^2$ separately.  The $\rho$ integral is of the familiar form $\int_{0}^{\infty}d\rho \, \rho^{\alpha-1}e^{-\rho}$ and can be done easily
\begin{equation}
 \begin{aligned}
W_{2,0}(P_1,P_2) & =\frac{\pi^{\frac{d-q}{2}}\Gamma\big(\frac{\Delta_1+\Delta_2+q-d}{2}\big)}{2 \, \Gamma(\Delta_1)\Gamma(\Delta_2)}\int\frac{d\tau_1d\tau_2  d\alpha_1 d\alpha_2}{\alpha_1\alpha_2}\,\delta(1-\alpha_1-\alpha_2)\nonumber
\\[5pt]
&\times \frac{\Gamma\left(\frac{\Delta_1+\Delta_2-2\tau_1-2\tau_2}{2}\right)\Gamma(\tau_1)\Gamma(\tau_2)\alpha_1^{\Delta_1-\tau_1-\tau_2}\alpha_2^{\Delta_2-\tau_1-\tau_2}}{(x_1-x_2)^{2\tau_1}(2 x_{1,\perp}\cdot x_{2,\perp})^{\tau_2}(\alpha_1^2x_{1,\perp}^2+\alpha_2^2x_{2,\perp}^2)^{\frac{\Delta_1+\Delta_2-2\tau_1-2\tau_2}{2}}} \, .
\end{aligned}
\end{equation}
The integral over the Schwinger parameters evaluates to simple powers\footnote{This is done using a slight generalization of  Cheng Wu theorem \cite{Smirnov:2012gma} which states that the integral with  $\beta\nu=\sum_{i}\Delta_i$
\begin{align}
\int \prod_{i}\frac{\alpha_i^{\Delta_i}d\alpha_i}{\alpha_i}\,\frac{\delta(1-\sum_{i}\alpha_i)}{(\sum_{i}\alpha_i^{\beta}c_i)^{\nu}} =  \int \prod_{i}\frac{\alpha_i^{\Delta_i}d\alpha_i}{\alpha_i}\,\frac{\delta(1-H(\alpha_i))}{(\sum_{i}\alpha_i^{\beta}c_i)^{\nu}} \, ,
\end{align}
where $H(\alpha_i)$ is any hyperplane defined using the variables $\alpha_i$. This statement can be easily proved by using a Schwinger representation for the denominator $(\sum_{i}\alpha_i^{\beta}c_i)^{\nu}$ and then rescaling all the variables $\alpha_i$ by this new Schwinger parameter. The condition $\beta\nu=\sum_{i}\Delta_i$ guarantees that the the dependence on this new parameter will only show up in the delta function. The last step is to notice that we can change the delta function by any other that just contains an hyperplane $H(\alpha_i)$.  A simple choice is to let $H(\alpha_i)=\alpha_1$, which effectively sets the variable $\alpha_1=1$. Then we can change variables $\alpha_i\rightarrow \alpha_i^{1/\beta}$ and arrive at an expression that can more easily be recognized as the one of the Feynman parametrization of simple powers. 
}
\begin{align}
W_{2,0}=&\frac{\pi^{\frac{d-q}{2}}\Gamma\big(\frac{\Delta_1+\Delta_2+q-d}{2}\big)}{4 \, \Gamma(\Delta_1)\Gamma(\Delta_2)(x_{1,\perp}^2)^{\frac{\Delta_1}{2}}(x_{2,\perp}^2)^{\frac{\Delta_2}{2}}}\nonumber\\
&\times\int d\delta_{12}d\nu_{12} \,\xi^{-\delta_{12}}(2\cos\phi)^{-\nu_{12}}\Gamma(\delta_{12})\Gamma(\nu_{12})\prod_i\Gamma\left(\frac{\Delta_i-\delta_{12}-\nu_{12}}{2}\right).
\end{align}
Notice that the factor inside the last two $\Gamma$ functions is precisely the variable $\alpha_i/2$. It appears that this Witten diagram is expressed, as expected, with two cross ratios $\xi$ and $\cos \phi$. However this is not actually correct since this particular Witten diagram can be expressed in terms of the cross ratio $\chi$ defined previously. The result expressed in terms of this variable is much simpler and is given by
\begin{align}
W_{2,0}=\frac{\pi^{\frac{d-q}{2}}\Gamma\big(\frac{\Delta_1+\Delta_2+q-d}{2}\big)}{4 \, \Gamma(\Delta_1)\Gamma(\Delta_2)(x_{1,\perp}^2)^{\frac{\Delta_1}{2}}(x_{2,\perp}^2)^{\frac{\Delta_2}{2}}}\int d\tau \chi^{-\tau}\Gamma(\tau)\prod_i\Gamma\left(\frac{\Delta_i-\tau}{2}\right).
\end{align}
This distinct feature of dCFT as compared to CFT, since in the later contact interactions without derivatives in AdS give rise to functions depending on two cross ratios. However we expect an explicit dependence on the other cross ratio $\cos \phi$ whenever one considers contact interactions with derivatives. The methods developed in \cite{Costa:2014kfa} might be useful to study this type of interactions.

The case with $n$  bulk and $m$ defect operators  can be obtained with the same steps explained above and in the boundary case \cite{Rastelli:2017ecj}. This straightforward exercise leads to 
\begin{equation}
 \begin{aligned}
W_{n,m}=\mathcal{N}\int &[d\beta_{IJ}][d\tau_{ij}][d\gamma_{iI}]\,(-2P_i \gbullet P_j)^{-\tau_{ij}} P_{iI}^{-\gamma_{iI}}\hat{P}_{IJ}^{-2\beta_{IJ}}(P_i\circ P_i)^{-\frac{\alpha_i}{2}}
\\[5pt]
&\times\Gamma(\gamma_{ij})\Gamma(\beta_{ij})\Gamma(\theta_{iI})\Gamma(\nu_{IJ})\Gamma\big(\frac{\alpha_i}{2}\big) \, ,
\end{aligned}
\end{equation}
where
\begin{align}
\mathcal{N}=\frac{\pi^{\frac{d-q}{2}}\Gamma\left(\frac{\sum_i\Delta_i+\sum_I\hat{\Delta}_I-(d-q)}{2}\right)}{2^{n+1}\prod_{i}\Gamma(\Delta_i)\prod_{I}\Gamma(\hat{\Delta}_I)} \, .
\end{align}
We have decided to use the product $P_i \gbullet P_j$ instead of $P_i \cdot P_j$ and $P_i\circ P_j$ for simplicity. The result with $P_i \cdot P_j$ and $P_i\circ P_j$ can be done without any problem.  

\subsection*{Exchange Witten diagrams - bulk case}
Exchange Witten diagrams have been recently computed for boundary CFTs. The results share many similarities with the CFT analogue. For instance, it is easier to obtain a closed formula in the case where the external and exchanged dimensions  $(\Delta_1+\Delta_2-\Delta)/2$ is a non negative integer number and both can be solved using the split representation for the bulk to bulk propagator. 

The generalization of an exchange Witten diagram for the defect setting with generic co-dimension is straightforward. So we will just review here the main steps involved in \cite{Rastelli:2017ecj}. The Witten diagram that we consider is given by the integral, 
\begin{align}
W_{\textrm{bulk}}= \int_{AdS_{d+1-q}} dW \int_{AdS_{d+1}} dX G_{B\partial}^{\Delta_1}(P_1,X)G_{B\partial}^{\Delta_2}(P_2,X) G_{BB}^{\Delta}(X,W).
\end{align}
  The special case of $(\Delta_1+\Delta_2-\Delta)/2$ can be done using the trick explained in \cite{Howtozintegrals,Rastelli:2017ecj}.  First one computes the $X$ integral 
\begin{align}
A(W,P_1,P_2)=  \int_{AdS_{d+1}} dX G_{B\partial}^{\Delta_1}(P_1,X)G_{B\partial}^{\Delta_2}(P_2,X) G_{BB}^{\Delta}(X,W) \, ,
\end{align}
which can be expressed in terms of an hypergeometric function. For dimensions satisfying $(\Delta_1+\Delta_2-\Delta)/2 \in Z$ the hypergeometric simplifies into a finite sum of terms and consequently the exchanged diagram turns into 
\begin{align}
A(W,P_1,P_2)= \sum_{k=k_{min}}^{k_{max}}a_k \frac{G_{B\partial}^{k+\Delta_1-\Delta_2}(P_1,W)G_{B,\partial}^{k}(P_2,W)}{P_{12}^{\Delta_2-k}} \, ,
\end{align}
where 
\begin{equation}
 \begin{aligned}
& a_{k-1}=a_k\frac{(k-\frac{\Delta}{2}+\frac{\Delta_{12}}{2})(k-\frac{d}{2}+\frac{\Delta}{2}+\frac{\Delta_{12}}{2})}{(k-1)(k-1+\Delta_{21})} \, , \qquad a_{\Delta_2-1}=\frac{1}{4(\Delta_1-1)(\Delta_2-1)} \, ,
\\[5pt]
& k_{min}=\frac{\Delta-\Delta_1+\Delta_2}{2} \, , \qquad\qquad k_{\textrm{max}}=\Delta_2-1 \, .
\end{aligned}
\end{equation}
Notice that the remaining integrals resume to the case of the two-point contact Witten diagram. Then, it follows that the correlator for this diagram is given by
\begin{align}
W_{\textrm{bulk}}&= \sum_{k_{\textrm{min}}}^{k_{\textrm{max}}}\frac{a_k}{P_{12}^{\Delta_2}}\frac{\pi^{\frac{d-q}{2}}\Gamma(k+\Delta_{12}+k+q-d)\xi^{k-\Delta_2}}{\Gamma(k+\Delta_{12})\Gamma(k)(P_1\circ P_1)^{\frac{\Delta_{1}}{2}}(P_2\circ P_2)^{\frac{\Delta_{2}}{2}}}\nonumber\\
&\times\int d\tau \,\Gamma(\tau)\Gamma(k-\tau)\Gamma(k+\Delta_{12}-\tau) \chi^{-\tau} \, .
\end{align}
We have obtained another correlator that only depends on one cross ratio $\chi$. A non-trivial dependence on another cross ratio should come from interaction terms with derivatives. 

\subsection*{Exchange Witten diagrams - bulk case with generic dimensions}
The generalization for generic dimensions follows the case of the bCFT\cite{Rastelli:2017ecj}. First one writes the bulk to bulk propagator as two bulk-to-boundary integrated over a boundary point $P$\cite{Penedones:2010ue}
\begin{align}
G_{BB}^{\Delta}= \int_{-i\infty}^{i\infty}\frac{dc}{(\Delta-h)^2-c^2}\frac{\Gamma(h+c)\Gamma(h-c)}{\Gamma(c)\Gamma(-c)2\pi^d}\int_{\partial AdS} \frac{dP_0}{(-2P_0\cdot X)^{h+c}(-2P_0\cdot W)^{h-c}}  \, ,
\end{align} 
where $h = d/2$ and the integration over $c$ run parallel to the imaginary axis. This representation for the bulk to bulk propagator has the advantage that it makes easy the integration over $AdS$
\begin{align}
W_{bulk}= \int_{-i\infty}^{i\infty}\frac{dc}{(\Delta-h)^2-c^2}\int_{\partial}\frac{dP_0 \,\Phi_h(c)}{P_{12}^{\frac{\Delta_1+\Delta_2-c-h}{2}}P_{01}^{\frac{\Delta_{12}+h+c}{2}}P_{02}^{\frac{\Delta_{21}+h+c}{2}}}\frac{1}{(P_0\circ P_0)^{\frac{h-c}{2}}} \, ,
\end{align}
with $\Phi_h(c)$ defined by
\begin{align}
\Phi_h(c) =\frac{\Gamma(h+c)\Gamma(h-c)}{\Gamma(c)\Gamma(-c)2\pi^d},\,  \ \ \Delta_{ij}=\Delta_i-\Delta_j. 
\end{align}
The integration over the boundary of $AdS$ is done by writing the propagators using Schwinger parameters
\begin{align}
&\int\frac{dP_0}{P_{01}^{a_1}P_{02}^{a_2}}\frac{1}{(P_0\circ P_0)^{a_3}}= \frac{\pi^{d/2}}{\Gamma(a_1)\Gamma(a_2)\Gamma(a_3)}\int\frac{d\alpha_1d\alpha_2d\alpha_3}{\alpha_1\alpha_2\alpha_3}\alpha_1^{a_1}\alpha_2^{a_2}\alpha_3^{a_3}\nonumber\\
&\times \frac{e^{-\frac{\alpha_1\alpha_2(x_1-x_2)^2}{\alpha_1+\alpha_2}-\frac{\alpha_3(\alpha_1^2x_{1,\perp}^2+\alpha_2^2x_{2,\perp}^2+2\alpha_1\alpha_2x_{1,\perp}\cdot x_{2,\perp})}{(\alpha_1+\alpha_2)(\alpha_1+\alpha_2+\alpha_3)}}}{(\alpha_1+\alpha_2)^{\frac{d-q}{2}}(\alpha_1+\alpha_2+\alpha_3)^{q/2}}.
\end{align} 
The rest of the computation that follows is the usual: a) we introduce the identity in the form of two integrated $\delta$-functions, 
\begin{align}
1=\int d\lambda \, \delta \big(\lambda-(\alpha_1+\alpha_2+\alpha_3) \big)\int d\rho \, \delta \big(\rho-(\alpha_1+\alpha_2) \big) \, ,
\end{align}
b) we rescale $\alpha_1,\alpha_2\rightarrow \rho \, \alpha_i$ and c) we use an inverse Mellin transform on the factors $(x_1-x_2)^2$ and $2x_{1,\perp}\cdot x_{2,\perp}$. After that we find
\begin{align}
& \frac{\pi^{d/2}}{\Gamma(a_1)\Gamma(a_2)\Gamma(a_3)}\int\frac{d\alpha_1d\alpha_2d\alpha_3 d\lambda d\tau_1 d\tau_2}{\alpha_1\alpha_2\alpha_3}\alpha_1^{h+\tau_1-a_2-\alpha_3}\alpha_2^{h+\tau_1-a_1-a_3}\alpha_3^{a_3-\tau_1-\tau_2} \nonumber\\
&\times  \frac{\delta(1-\alpha_1-\alpha_2)\delta(\lambda-1-\alpha_3)\lambda^{\tau_1+\tau_2-q/2}\Gamma(a_1+a_2+a_3-h-\tau_1-\tau_2)\Gamma(\tau_1)\Gamma(\tau_2)}{(2x_{1,\perp}\cdot x_{2,\perp})^{\tau_2}(\alpha_1^2x_{1,\perp}^2+\alpha_2^2x_{2,\perp}^2 )^{\tau_1} \big( (x_1-x_2)^2 \big)^{a_1+a_2+a_3-h-\tau_1-\tau_2}} \, .
\end{align}
Notice that the $\alpha_1,\alpha_2$ integrals evaluated to simple powers since $(h+\tau_1-a_2-\alpha_3)+(h+\tau_1-a_1-a_3)=2\tau_1$. The other two integrals, in $\lambda$ and $\alpha_3$, can be evaluated straightforwardly giving 
\begin{align}
&\int  \frac{d\tau_1 d\tau_2\Gamma(\frac{\Delta_{12}+2\tau_1}{2})\Gamma(\frac{\Delta_{21}+2\tau_1}{2})}{\Gamma(2\tau_1)(x_{1,\perp}^2)^{\frac{\Delta_{12}+h+c-\tau_1}{2}}(x_{2,\perp}^2)^{\frac{\Delta_{21}+h+c-\tau_1}{2}}} \frac{\Gamma(\frac{c+h-2(\tau_1+\tau_2)}{2})\Gamma(\tau_1)\Gamma(\tau_2)\Gamma(\frac{c-h+q}{2})\Gamma(\frac{h-c-2(\tau_1+\tau_2)}{2})}{(2x_{1,\perp}\cdot x_{2,\perp})^{\tau_2}P_{12}^{\frac{c+h-2(\tau_1+\tau_2)}{2}}\Gamma \big( \frac{q}{2}-\tau_1-\tau_2 \big)}\nonumber.
\end{align}
The Witten diagram is obtained by putting all pieces together
\begin{align}
W_{\textrm{bulk}}=&\int_{-i\infty}^{i\infty}\frac{dc}{(\Delta-h)^2-c^2}\frac{\Phi_h(c)}{P_{12}^{\frac{\Delta_1+\Delta_2-c-h}{2}}}\int  \frac{d\tau_1 d\tau_2\Gamma(\frac{\Delta_{12}+2\tau_1}{2})\Gamma(\frac{\Delta_{21}+2\tau_1}{2})}{\Gamma(2\tau_1)(x_{1,\perp}^2)^{\frac{\Delta_{12}+h+c-\tau_1}{2}}(x_{2,\perp}^2)^{\frac{\Delta_{21}+h+c-\tau_1}{2}}}
\nonumber\\[5pt]
&\times \frac{\Gamma\Big( \frac{c+h-2(\tau_1+\tau_2)}{2} \Big)\Gamma(\tau_1)\Gamma(\tau_2)\Gamma(\frac{c-h+q}{2})\Gamma \Big(\frac{h-c-2(\tau_1+\tau_2)}{2} \Big)}{(2x_{1,\perp}\cdot x_{2,\perp})^{\tau_2}P_{12}^{\frac{c+h-2(\tau_1+\tau_2)}{2}}\Gamma\big( \frac{q}{2}-\tau_1-\tau_2 \big)} \, .
\end{align}
We were not able to write this Witten diagram using just one cross ratio. The simplification in the previous case was specific to the condition on the dimension of the operators that were involved. 

\subsection*{Exchange Witten diagrams - defect case with generic dimensions}
Defect exchange Witten diagrams can be obtained in a similar way using the spectral representation for the bulk to bulk propagator on the defect
\begin{align}
&W_{\textrm{defect}}= \int_{AdS_{d+1-q}}\!\!\!\!\!\!dW_1\int_{AdS_{d+1-q}}\!\!\!\!\!\!dW_2\, G_{B\partial}^{\Delta_1}(P_1,W_1)G_{BB}^{\Delta}(W_1,W_2)G_{B\partial}^{\Delta_2}(P_2,W_2),\,
\\[5pt]
&G_{BB}^{\Delta}=\int_{-i\infty}^{i\infty}\frac{dc}{(\Delta-h')^2-c^2}\Phi_{h'}(c)\int \frac{d\hat{P}}{(-2\hat{P}\cdot W_1)^{h'+c}(-2\hat{P}\cdot W_2)^{h'-c}} \, , \qquad h'=\frac{d-q}{2}.\nonumber
\end{align}
The integration over $AdS$ is the same as the one already done for the contact diagram $W_{1,1}$. At the end of the day there is only the defect integral to compute
\begin{align}
&\int \frac{d^{d-q}\bar{p}}{(x_{1,\perp}^2+(\bar{x}_1-\bar{p})^2)^{h'+c}(x_{2,\perp}^2+(\bar{x}_2-\bar{p})^2)^{h'-c}}\\
&=\frac{\pi^{h'}}{\Gamma(h'+c)\Gamma(h'-c)}\int\frac{d\alpha_1d\alpha_2d\rho}{\alpha_1\alpha_2\rho}\delta(1-\alpha_1-\alpha_2)\rho^{h'}\alpha_1^{h'+c}\alpha_2^{h'-c}e^{-\rho \left( \alpha_1x_{1,\perp}^2+\alpha_2x_{2,\perp}^2+\alpha_1\alpha_2(\bar{x}_1-\bar{x}_2)^2 \right)} \, ,\nonumber
\end{align}
where we have introduced Feynman parameters and a $\delta$-function as in the previous computations. Now we can use the inverse Mellin representation for the exponential of the last term.  
The result for a defect exchange Witten diagram is 
\begin{align}
W_{\textrm{defect}} =& \int_{-i\infty}^{i\infty}\frac{dc}{(\Delta-h')^2-c^2}\frac{\Phi_{h'}(c) \, \pi^{h'}}{\Gamma(h'+c)\Gamma(h'-c)}\times\nonumber\\
&\int d\tau \frac{\Gamma(\tau)\Gamma(h'+\tau)\Gamma(\tau+h'+c)\Gamma(\tau+h'-c)}{\Gamma(2\tau+h')}\left(\frac{\hat{P}_{12}}{x_{\perp,1}^2x_{\perp,2}^2}\right)^{\tau} \, .
\end{align} 
All examples of Witten diagrams that were considered in this paper had simple contact interactions. It would be interesting to include interactions with derivatives to probe the full structure of the defect.

\section{Discussion}
In this paper we have analyzed in more detail the analytic structure of Mellin amplitudes for defect conformal field theories. Recall that the analytic structure played an important role in the Mellin bootstrap papers \cite{Rastelli:2016nze,Rastelli:2017ymc,Zhou:2017zaw,Rastelli:2017udc,Gopakumar:2016wkt,Gopakumar:2016cpb,Dey:2017oim,Dey:2016mcs}. Thus, one future direction from this work is to try to extend the Mellin bootstrap to a defect conformal field theory. Here the boundary/interface setting is a good starting point since the two-point function depends on just one variable but still it has interesting physics. 

Another problem that is relative is the systematic study of higher point functions. We have study the location of the poles of the Mellin amplitudes however it still remains to be  studied the factorization properties of Mellin amplitudes with more than two bulk operators. This problem can be probably solved with the same methods that were applied for the CFT case\cite{Goncalves:2014rfa}. In order to derive the factorization properties in full generality it would also be interesting to study Mellin amplitudes for operators with spin. 

In \cite{Rastelli:2017ecj} it was shown that boundary conformal blocks can be written as geodesic Witten diagrams\cite{Hijano:2015zsa}. The same arguments that lead to this conclusion for the co-dimension one defect\cite{Rastelli:2017ecj} can be equally applied to this more general case.

Recently \cite{Lemos:2017vnx} it was shown that in every defect conformal field theory with co-dimension higher than one has defect operators whose dimension goes asymptotically to $\hat{\Delta}\approx s+\Delta_{\phi}+s+2m,\,\,\,s\rightarrow \infty$. One possible direction is to derive the same results and corrections in Mellin space. 

It has been possible to find a correspondence between the number of solutions of bootstrap equations with a sparse spectrum and AdS contact interactions for a CFT \cite{JP}. It is tempting to see if the same could be done for a dCFT. This would probably require the study of Witten diagrams with contact interactions with derivatives which were not studied in this paper.  

The flat space limit can be written in a simple form using Mellin amplitudes. In \cite{JPMellin} Penedones derived a relation between  Mellin amplitudes of contact interactions in AdS and the corresponding flat space scattering amplitude. The same idea could be tried for the Mellin amplitude of the dCFT. 

\section*{Acknowledgments}
We thank Joao Penedones for useful discussions. We also thank Edoardo Vescovi and Marco Meineri for reading the draft and their useful comments. 

V.G. is funded by FAPESP grant 2015/14796-7. The work of GI is supported by FAPESP grant 2016/08972-0 and 2014/18634-9.

\appendix
\section{Embedding space toolbox}
The embedding space formalism is useful to study conformal field theories with or without the presence of defects since the conformal transformations become linear in the embedding space. In this appendix we collect some results to make it easier for the reader to translate the correlators to the more familiar position space\footnote{The interested reader can find more details in \cite{Billo:2016cpy,Lauria:2017wav}.}. 

The coordinates are split in two sets such that the action of the groups $SO(p+1,1)$ and $SO(q)$ on them is linear. The first set correspond to parallel directions while the second to orthogonal to the defect. The indices are also split into two sets with the parallel being label with $A,B\dots$ and the orthogonal with $I,J,\dots$. 

We use two scalar products with respect to each of the symmetry groups
\begin{align}
P_1 \gbullet P_2 = P_1^{A}P_2^{B}\eta_{AB} \, , \qquad P_1\circ P_2 = P_1^{I}P_2^{J}\delta_{IJ} \, ,
\end{align}
where the indices $A$ and $B$ run from $1$ to $2+p$ and $I$ and $J$ from $1$ to $q$. Let us now list the projection to physical space of both these scalar products
\begin{align}
-2P_1 \gbullet P_2= |x_{12}^{a}|^2+|x_{1}^{i}|^2+|x_{2}^{i}|^2 \, \qquad P_1\circ P_2=x_{1}^{i}x_{2}^{i} \, ,
\end{align}
where $a$ are physical space parallel coordinates to the defect while $i$ corresponds to orthogonal.  

The $SO(d+1,1)$ embedding scalar product, usually denoted by a simple dot, 
\begin{align}
-2P_1\cdot P_2 = (x_1-x_2)^2
\end{align}
can be written in terms of the parallel and orthogonal scalar products.

\bibliographystyle{./utphys}
\bibliography{./mybib}

\end{document}